\documentclass[runningheads,a4paper]{llncs}

\usepackage{amssymb}
\usepackage{graphicx}
\usepackage{url}
\usepackage{hyperref}
\usepackage{apacite}
\newcommand{\keywords}[1]{\par\addvspace\baselineskip
\noindent\keywordname\enspace\ignorespaces#1}

\begin{document}


\mainmatter  

\newcommand{\TT}[1]{$\texttt{#1}$}
\newcommand{\T}[1]{\texttt{#1}}
\newcommand{\systemName}{MMM}

\title{MMM : Exploring Conditional Multi-Track Music Generation with the Transformer}

\titlerunning{MMM : Exploring Conditional Multi-Track Music Generation}

%
%
\author{Jeff Ens \and Philippe Pasquier \thanks{We acknowledge the support of the Natural Sciences and Engineering Research Council of Canada (NSERC), and the Helmut \& Hugo Eppich Family Graduate Scholarship}}
%
\authorrunning{Jeff Ens and Philippe Pasquier}

\institute{Simon Fraser University\\ \email{jeffe@sfu.ca}}

\maketitle

\begin{abstract}

We propose the Multi-Track Music Machine (MMM), a generative system based on the Transformer architecture that is capable of generating multi-track music. In contrast to previous work, which represents musical material as a single time-ordered sequence, where the musical events corresponding to different tracks are interleaved, we create a time-ordered sequence of musical events for each track and concatenate several tracks into a single sequence. This takes advantage of the Transformer's attention-mechanism, which can adeptly handle long-term dependencies. We explore how various representations can offer the user a high degree of control at generation time, providing an interactive demo that accommodates track-level and bar-level inpainting, and offers control over track instrumentation and note density.  

\end{abstract}

\section{Introduction}




Research involving generative music systems has focused on modelling musical material as an end-goal, rather than on the affordances of such systems in practical scenarios \cite{sturm2019machine}. As a result, there has been a focus on developing novel architectures and demonstrating that music generated with these architectures is of comparable quality to human-composed music, often via a listening test. Although this is a necessary first step, as systems must be capable of generating compelling material before they can be useful in a practical context, given the impressive capabilities of the Transformer-based models in the music domain \cite{donahue2019lakhnes,huang2018music}, we shift our focus to increasing the affordances of a Transformer-based system. To achieve this goal, we develop a novel representation for multi-track musical material that accommodates a variety of methods for generation.

To avoid any confusion, we will first define what constitutes multi-track music. We consider a {\it track} to be a collection of notes played on a single instrument. Although alternate terminology has been employed to describe tracks, which may be referred to as voices or instruments in various contexts, we believe that the term track is clearest, as there is a clear analog to the tracks present in a digital audio workstation. We avoid using the term voice, as it commonly connotes a monophonic musical line, while we wish to refer to musical material that may contain multiple notes sounding simultaneously. Furthermore, within a piece there may be multiple tracks featuring the same instrument, each playing a different musical part, which would make usage of the term instrument problematic. Consequently, multi-track music refers to material containing two or more tracks, where each track is played by a single instrument and may optionally contain multiple notes that sound simultaneously. It is also important to note the difference between {\it polyphonic tracks}, which contain simultaneously sounding notes, and {\it monophonic tracks}, which contain a single sequence of non-overlapping notes.

Given our interest in enhancing the usability of a system at generation time, it is worth reviewing different methods for generation, which we group into four categories: unconditioned, continuation, inpainting, and attribute-control. Unconditioned generation is analogous to generating music from scratch. Besides changing the data that the model is trained on, the user has limited control over the output of the model. Continuation involves conditioning the model with musical material that precedes (temporally) the music that is to be generated. Since both unconditioned generation and continuation come for free with any auto-regressive model trained on a temporally ordered sequence of musical events, most systems are capable of generating musical material in this manner. Inpainting conditions generation on a subset of musical material, asking the model to fill in the blanks, so to speak. Note that inpainting can occur at different levels (i.e. note-level, bar-level, track-level). CoCoNet \cite{huang2019counterpoint} allows for inpainting of Bach chorales on the bar and track level, while InpaintNet \cite{pati2019learning} allows for inpainting of 2-8 bars of monophonic musical material. Attribute-control involves conditioning generation on high-level attributes such as style, tempo or density. For example, music generated by MuseNet \cite{payne2019} can be conditioned on a set of instruments and a musical style. In some circumstances, generation methods can be chained, resulting in an iterative generation process. For example, a musical segment exhibiting a particular style could be generated via attribute control, and the user could then select various sections they are unsatisfied with for inpainting.

Our primary contribution is a novel representation for musical material, which when coupled with state-of-the-art transformer architectures, results in a powerful and expressive generative system. In contrast to previous work, which represents musical material as a single time-ordered sequence, where the musical events corresponding to different tracks are interleaved, we create a time-ordered sequence of musical events for each track and concatenate several tracks into a single sequence. Although the difference is subtle, this enables track-level inpainting, and attribute control over each track. We also explore variations on this representation which allow for bar-level inpainting. Unto our knowledge, both inpainting and attribute control have not been integrated into a single model.


\section{Related Work}

There are two main ways in which musical material is represented: as a matrix (i.e. a pianoroll), or as a sequence of tokens. A piano roll is a boolean matrix $x \in \{0,1\}^{T \times P}$, where $T$ is the number of time-steps and $P$ is the number of pitches. Typically $P = 128$, allowing the piano roll to represent all possible MIDI pitches, however, it is not uncommon to reduce the range of pitches represented \cite{dong2018musegan}. Multi-track musical material can be represented using a boolean tensor $x \in \{0,1\}^{M \times T \times P}$, where $M$ is the number of tracks. However, using this type of representation is inherently inefficient, as the number number of inputs increases by $T \times P$ for each track that is added, and accommodating small note lengths (ex. $32^{nd}$ triplets) substantially increases $T$. Despite these drawbacks, this representation has been used in practice \cite{boulanger2012modeling,dong2018musegan,huang2019counterpoint}. The alternative approach, is to represent musical material as a sequence of tokens, where each token corresponds to a specific musical event or piece of metadata. For example, the PerformanceRNN \cite{oore2018time} and Music Transformer \cite{huang2018music} use a token based representation comprised of 128 distinct \TT{NOTE\_ON} tokens, which are used to indicate the onset of a particular pitch; 128 \TT{NOTE\_OFF} tokens, which denote the end of a particular pitch; and 100 \TT{TIME\_SHIFT} tokens, which correspond to different time-shifts ranging from 10ms to 1 second. Although this type of representation can accommodate polyphony, it does not distinguish between different tracks or instruments.

Our work is most similar to LahkNES \cite{donahue2019lakhnes}, MusicVAE \cite{roberts2018hierarchical} and MuseNet \cite{payne2019}, which all employ token-based representations to model multi-track music. LahkNES models Nintendo Entertainment System (NES) data, which is comprised of 3 monophonic tracks and a drum track, using a transformer architecture. MusicVAE is trained on bass, melody and drum trios extracted from the Lahk MIDI Dataset (LMD) \cite{raffel2016learning}, which allows generation to be conditioned on a latent vector. MuseNet is trained on a superset of the LMD, and accommodates 10 different track types ranging from piano to guitar. Note that MuseNet supports both polyphonic and monophonic tracks. 




However, in contrast to these methods, where the musical events from several tracks are interleaved into a single time-ordered sequence, we concatenate single-instrument tracks, which allows for greater flexibility in several areas. First of all, we can decouple track information from \TT{NOTE\_ON} and \TT{NOTE\_OFF} tokens, allowing the use of the same \TT{NOTE\_ON} and \TT{NOTE\_OFF} tokens in each track. This differs from LahkNES, MusicVAE, and MuseNet, which use separate \TT{NOTE\_ON} and \TT{NOTE\_OFF} tokens for each track, placing inherent limitations on the number of tracks that can be represented. Even MuseNet, which is the largest of these networks, can only accommodate 10 different tracks. Secondly, using our representation, we are able to accommodate a wide variety of instruments, including all 128 general midi instruments, and a large number of tracks, without a employing a prohibitively large token vocabulary. In contrast to LahkNES and MusicVAE which are designed for a fixed-schema of tracks, our system can handle an arbitrary set of tracks. MuseNet is similar in this regard, however, it only supports 10 distinct instruments. Although MuseNet permits attribute control over the instrument, allowing the user to specify the set of instruments that will be featured in the generated excerpt, this information is only treated as a strong recommendation to the model, and does not guarantee which instruments will actually be used. Our system allows for specific attribute control over the instrument for each track, with the guarantee that a particular instrument will be used. Third, we offer the user control over the note-density of each track, which is not accomodated with LahkNES, MusicVAE or MuseNet. Finally, we allow for track-level and bar-level inpainting, which is not possible using LahkNES, MusicVAE and MuseNet. Collectively, these improvements afford the end-user a high-degree of control over the generated material, which has previously been proposed as a critical area of research \cite{briot2017deep}.







\section{Motivation}

Although systems which generate high-quality music have been proposed in recent years \cite{huang2018music,payne2019,liang2016bachbot,sturm2017taking}, their usage in practical contexts is limited for two different reasons. First of all, most models place restrictions on the nature of the input. In most cases, there are limitations placed on the number and type of tracks \cite{roberts2018hierarchical,payne2019}. Secondly, the user is not afforded fine-grained control over the generation process, which is critical for a system to be useful in the context of computational assisted composition. Even MusicVAE \cite{roberts2018hierarchical}, which incorporates a latent model of musical space, allowing for interpolation between examples, does not afford fine-grained control of the individual tracks. For example, it is not possible to freeze the melody and generate a new drum-part and bassline. Although one-shot generation of musical material is impressive from a technical standpoint, it is not that useful in a practical context, as the user may wish to create subtle variations on a fixed piece of music.

In contrast to time-ordered sequences, where most of the important dependencies, such as the most recently played notes, are in the recent history, non-time-ordered sequences frequently feature important dependencies in the distant history. For example, in our representation, simultaneously sounding notes in different tracks are spread far apart. The use of non-time-ordered representations is directly motivated the nature of the transformer attention mechanism \cite{vaswani2017attention}. In contrast to Recurrent Neural Networks (RNN), which sharply distinguish nearby context (the most recent 50 tokens) from the distant history \cite{khandelwal2018sharp}, attention-based architectures allow for distant tokens to be directly attended to if they are relevant to the current prediction. Consequently, we do not pay a significant penalty for training models on non-time-ordered sequences, where important dependencies are predominantly in the distant history, provided the necessary tokens are within the attention window. This directly motivates the usage of non-time-ordered sequences, as they facilitate rich conditional generation.

\section{Proposed Representation}


\begin{figure}[t!]
    \centering
    \includegraphics[width=.95\textwidth]{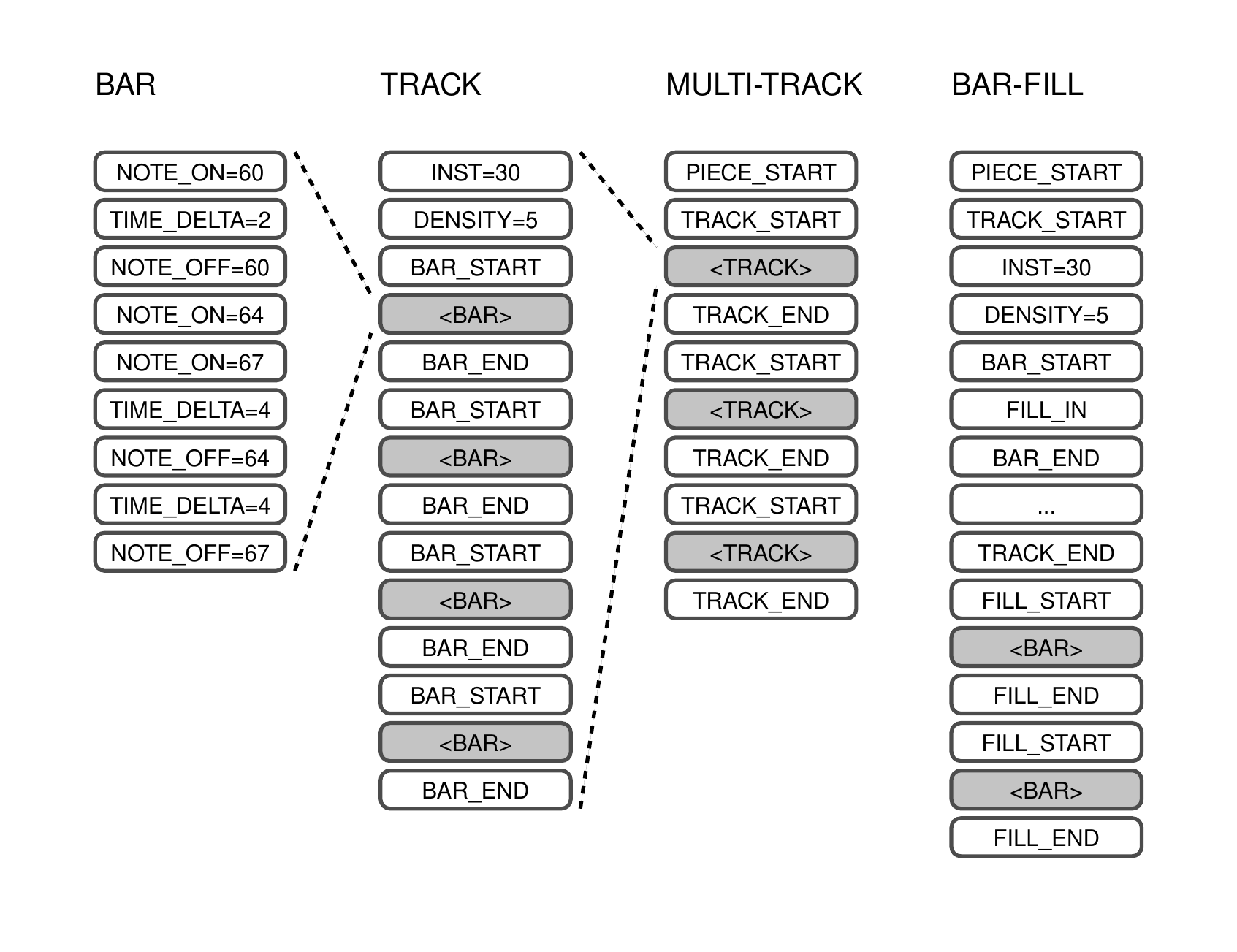}
    \caption{The MultiTrack and BarFill representations are shown. The \TT{<bar>} tokens correspond to complete bars, and the \TT{<track>} tokens correspond to complete tracks.}
    \label{fig:REP}
\end{figure}

To provide a comprehensive overview of the proposed representation, we first describe how a single bar of musical material is represented. Based on representations explored in previous studies \cite{oore2018time,huang2018music}, we represent musical material using 128 \TT{NOTE\_ON} tokens, 128 \TT{NOTE\_OFF} tokens, and 48 \TT{TIME\_SHIFT} tokens. Since musical events are quantized using 12 subdivisions per beat, 48 \TT{TIME\_SHIFT} tokens allow for the representation of any rhythmic unit from sixteenth note triplets to a full 4-beat bar of silence. Each bar begins with a \TT{BAR\_START} token, and ends with a \TT{BAR\_END} token. Tracks are simply a sequence of bars delimited by \TT{TRACK\_START} and \TT{TRACK\_END} tokens. At the start of each track, immediately following the \TT{TRACK\_START} token, an \TT{INSTRUMENT} token is used to specify the MIDI program which is to be used to play the notes on this particular track. Since there are 128 possible MIDI programs, we have 128 distinct \TT{INSTRUMENT} tokens. A \TT{DENSITY\_LEVEL} token follows the \TT{INSTRUMENT} token, and indicates the note density of the current track. A piece is simply a sequence of tracks, however, all tracks sound simultaneously rather than being played one after the other. A piece begins with the \TT{PIECE\_START} token. This process of nesting bars within a track and tracks within a piece is illustrated in Figure \ref{fig:REP}. Notably, we do not use a \TT{PIECE\_END} token, as we can simply sample until we reach the $n^{th}$ \TT{TRACK\_END} token if we wish to generate $n$ tracks. We refer to this representation as the MultiTrack representation.

Using the MultiTrack representation, the model learns to condition the generation of each track on the tracks which precede it. At generation time, this allows for a subset of the musical material to be fixed while generating additional tracks. However, while the MultiTrack representation offers control at the track level, it does not allow for control at the bar level, except in cases where the model is asked to complete the remaining bars of a track. Without some changes, it is not possible to generate the second bar in a track conditioned on the first, third, and fourth bars. In order to accommodate this scenario, we must guarantee that the bars on which we want to condition precede the bars we wish to predict, in the sequence of tokens that is passed to the model. To do this, we remove all the bars which are to be predicted from the piece, and replace each bar with a \TT{FILL\_PLACEHOLDER} token. Then, at the end of the piece (i.e. immediately after the last \TT{TRACK\_END} token), we insert each bar, delimiting each bar with \TT{FILL\_START} and \TT{FILL\_END} tokens instead of \TT{BAR\_START} and \TT{BAR\_END} tokens. Note that these bars must appear in the same order as the they appeared in the original MultiTrack representation. We refer to this representation as the BarFill representation. Note that the MultiTrack representation is simply a special case of the BarFill representation, where no bars are selected for inpainting.



\section{Training}



We use the Lahk MIDI Dataset (LMD) \cite{raffel2016learning}, which is comprised of 176,581 MIDI files. In order to explain how we derive token sequences from MIDI files, it is necessary to provide an overview of the MIDI protocol. There are three formats for MIDI files. Type 0 MIDI files are comprised of a header-chunk and a single track-chunk. Both Type 1 and 2 MIDI files contain a header-chunk and multiple track-chunks, however, the tracks in a Type 1 MIDI file are played simultaneously, while tracks in a Type 2 MIDI file are played sequentially. Since only $0.03\%$ of the LMD are Type 2 MIDI files, and the library we use for midi parsing does not support this encoding, we simply ignore them. Within a track-chunk, musical material is represented as a sequence of MIDI messages each which specify a channel and the time delta since the last message. In addition to note-on and note-off messages, which specify the onset and end of notes, patch-change messages specify changes in timbre by selecting one of 128 different instruments. To formally define a track, consider a Type 1 MIDI file $\T{F} = \{t^1,..,t^k\}$ comprised of $k$ track-chunks, where each track-chunk $t^i = \{m^i_i,..,m^i_{n_i}\}$ is an ordered set of $n_i$ MIDI messages. Note that a Type 0 MIDI file is simply a special case where $k=1$. Let $\T{chan}(x)$ (resp. $\T{inst}(x)$) be a function that returns the channel (resp. instrument) on which the message $x$ is played. Then, we can define a track as the set of midi messages $t_{i,c,k} = \{m_{\ell}^k : \T{inst}(m_{\ell}^k) = i, \T{chan}(m_{\ell}^k) = c, m_{\ell}^k \in t^k, t^k \in \T{F}\}$ that are found on the $k^{th}$ track-chunk, and played on the $c^{th}$ channel using the $i^{th}$ MIDI instrument. For example, given a MIDI file $\T{F} = \{t^1,t^2\}$, where $t^1 = \{m_1^1\}$, $t^2 = \{m^2_1,m^2_2\}$, $\T{chan}(m^1_1) = 0$, $\T{inst}(m^1_1) = 0$, $\T{chan}(m^2_1) = 3$, $\T{inst}(m^2_1) = 0$, $\T{chan}(m^2_2) = 3$, and $\T{inst}(m^2_2) = 34$, there would be three tracks $(t_{0,0,1}, t_{0,3,2}, t_{34,3,2})$.


For each of the 128 general MIDI instruments, we calculate the number of note onsets for each bar in the dataset, and use the quantiles of the resulting distributions to define distinct note-density bins for each MIDI instrument. Note that using the same note-density bins for all instrument types would be problematic as note-density varies significantly between instruments. We use 10 different note-density bins, where the $i^{th}$ bin is bounded by the $10i$ (lower) and $10(i+1)$ (upper) quantiles. We train a GPT2 \cite{radford2019language} model using the HuggingFace Transformers library \cite{Wolf2019HuggingFacesTS} with 8 attention heads, 6 layers, an embedding size of 512, and an attention window of 2048. We train two types of models: \systemName Bar, which is trained using the BarFill representation; \systemName Track, which is trained using the MultiTrack representation. We train 4-bar and 8-bar versions of \systemName Bar and \systemName Track. For 4-bar (resp. 8-bar) models we provide the model with at most 12 (resp. 6) tracks. Each time we select a $n$-bar segment, we randomly order the tracks so that the model learns each possible conditional between different types of tracks. When training the \systemName Bar models, we also select a random subset of bars for inpainting. 

\section{Using \systemName}

\if false
\begin{figure}
    \centering
    \includegraphics[width=\textwidth]{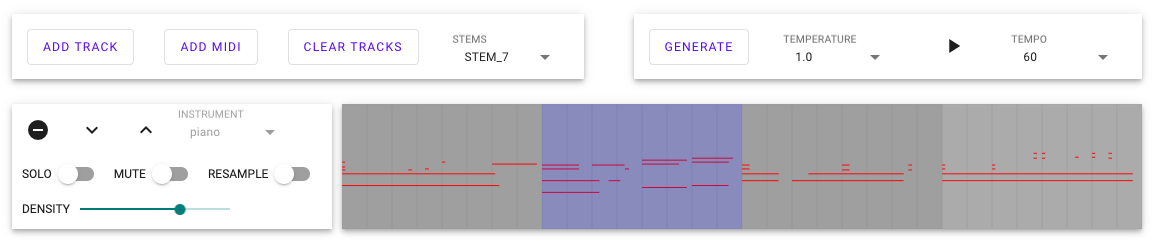}
    \caption{The user interface for the interactive demo. The bar shown in blue has been selected for bar-level inpainting.}
    \label{fig:GUI}
\end{figure}
\fi


In order to illustrate the flexibility of \systemName, we make available \footnote{\url{https://jeffreyjohnens.github.io/MMM/}} examples generated by the system, and an interactive demo. The demo was developed in Google Colab, making it accessible to all users with a compatible internet browser. The interface automatically selects the appropriate model, either \systemName Bar or \systemName Track, based on the bars or tracks that are selected for generation. We briefly outline the various ways that one can interact with \systemName\, when generating musical material.

\begin{enumerate}
    \item Track Inpainting : Given a possibly empty set of tracks $\mathbf{t} = \{t^1,...,t^k\}$, we can generate $n$ additional tracks. When the set of tracks is empty, this is equivalent to unconditioned generation. To do this, we condition the model with the tokens representing $k$ tracks and then sample until the $n^{th}$ \TT{TRACK\_END} token is reached.
    \item Bar Inpainting : Given a set of tracks $\mathbf{t} = \{t^1,...,t^k\}$ and a set of bars $\mathbf{b} = \{b_1,...,b_n\}$, we can resample each bar in $\mathbf{b}$. For this method, we condition the model with the tokens representing all the tracks, replacing each $b_i$ in $\mathbf{b}$ with the \TT{FILL\_PLACEHOLDER}. Then we sample until the $n^{th}$ \TT{FILL\_END} token is reached. 
    \item Attribute Control for Instruments: We can specify a set of MIDI instruments for each generated track, from which the model will choose. Practically, this is accomplished by masking the MIDI instruments we wish to avoid before sampling the \TT{INSTRUMENT} token at the start of a new track.
    \item Attribute Control for Note Density : We can specify the note density level for each generated track.
    \item Iterative Generation : The user can chain together various generation methods to iteratively compose a piece of music. Alternatively, generation methods can be chained automatically using a meta-algorithm. For example, given a set of tracks $\mathbf{t} = \{t^1,...,t^k\}$, we can progressively resample each track $t_i$ in $\mathbf{t}$ by asking the model to generate $(t^i | \{t^j : t^j \in \mathbf{t}, j \neq i\})$ for each $1 \leq i \leq k$. This bears some similarity to Gibbs sampling.  The resulting output should be more similar to the input than simply generating a set of tracks from scratch. Iterative generation also affords the user the opportunity to iteratively explore variations on generated material, or gradually refine a piece by progressively resampling bars which are not to their liking. 
\end{enumerate}


\section{Conclusion}

One current limitation, is that the model only allows for a fixed number of bars to be generated. Although approximately 99.8\% of 10-track 4-bar segments, 86.8\% of 10-track 8-bar segments and 38.8\% of 10-track 16-bar segments in the LMD can be represented using less than 2048 tokens, with some slight modifications to the architecture and representation, it should be possible to incorporate additional musical material. The transformer-XL architecture \cite{dai2019transformer} allows for extremely distant tokens to influence the current prediction via a hidden state, combining the strengths of the attention and recurrent mechanisms. Using this type of model, the current $n$-bar window could be conditioned on previous and future (if they are known) $n$-bar windows via the hidden state. Implementing additional types of attribute-control is an interesting area for future work. For example, conditioning generation on a particular genre or emotion would offer increased control at generation time. However, we must note that this type of control is available to a certain extent in the current model. Since \systemName\, offers conditional generation, the genre or emotion of the generated bars or tracks should reflect the genre or emotion of the content they are conditioned on. For example, if generation is conditioned on a jazz style drum track, generated tracks or bars should be consistent with this style. In addition, future work will include a more rigorous evaluation of the system itself. We have introduced a novel approach to representing musical material that offers increased control over the generated output. This offers a new and exciting avenue for future work, harnessing the strengths of the Transformer architecture to provide fine-grained control for the user at generation time.


\newpage

\bibliographystyle{apacite}
\bibliography{CSMC_MUME_LaTeX_Template.bib}

\end{document}